\documentclass[english, aps, prl, longbibliography, preprint]{revtex4-2}
\usepackage[T1]{fontenc}
\usepackage[latin9]{inputenc}
\usepackage{color}
\usepackage[english]{babel}
\usepackage{amssymb}
\usepackage{graphicx}
\usepackage{amsmath}
\usepackage{breakurl}
\usepackage{graphicx}
\usepackage{verbatim}
\usepackage[unicode=true,pdfusetitle, bookmarks=true,bookmarksnumbered=false,bookmarksopen=false, breaklinks=false,pdfborder={0 0 1},backref=false,colorlinks=true]{hyperref}
\hypersetup{linkcolor=blue, citecolor=blue, urlcolor  = blue}
\graphicspath{ {./figures/} } 


\begin{document}

\title{Size distribution of decaying foam bubbles}
\author{Ildoo Kim}
\email{ildookim@kku.ac.kr}
\affiliation{Department of Mechatronics, Konkuk University, Chungju, South Korea 27478}
\date{\today}

\begin{abstract}
The most studies on the stability of foam bubbles investigated the mechanical stability of thin films between bubbles due to the drainage by gravity.
In the current work, we take an alternative approach by assuming the rupture of bubbles as a series of random events and by investigating the time evolution of the size distribution of foam bubbles over a long time up to several hours. 
For this purpose, we first prepared layers of bubbles on Petri dishes by shaking soap solutions of a few different concentrations, and then we monitored the Petri dishes by using a time-lapse video imaging technique. 
We analyzed the captured images by custom software to count the bubble size distribution with respect to the initial concentration and elapsed time.
From the statistics on our data, we find that the total bubble volume decreases exponentially in time, and the exponent, i.e. the mean lifetime, is a function of the bubble size. 
The mean lifetimes of larger bubbles are observed to be shorter than those of smaller bubbles, by approximately a factor of 2.
\end{abstract}

\maketitle

\newpage

\section{Introduction}\label{sec1}

The stability of foam and bubbles is an intriguing topic that is relevant to many real-life applications in various industries, including cosmetics, mining, culinary arts, etc.
For several decade, this stability problem has attracted the attention of many scientists from diverse perspectives. \cite{Eri:2007tk, Zenit:2018ey, Miguet_2021, Yanagisawa_2021, Roux_2022}

In many previous studies, the topic has been approached from a mechanical perspective.
The foam consists of thousands of tiny bubbles that are bound to each other, forming soap films as walls between them.
As long as these walls are stable and intact, the foam structure is maintained; however, when they lose stability, the foam collapses. \cite{Isenberg_1992, Weaire_1999, Young_2011}
In this perspective, the drainage of the soap film under gravity becomes the main issue.
While the gravity-driven thinning of liquid film dates back to Newton's time, the recent studies focus on the drainage problem in conjunction with the viscoelasticity.
In general, the soap film's elasticity is defined as the change in the surface tension per the fractional change of area, that is,
\begin{equation}
    E=\frac{d\sigma}{d\ln A}= - h \frac{d\sigma}{d\ln h},
\end{equation}
where $A$ is the surface area of the film and $h$ is the thickness of the film.
A previous study by Sonin {\it et al.} reported that the stability of the film is influenced by the elasticity of the film \cite{Sonin:1993vj} although the relation is not linear. 
Later, a study by Georgieva {\it et al.} \cite{Georgieva:2009}, experimentally finds that high frequency elasitcity (Marangoni elasticity) involves thinning of the soap film, resulting in slower drainage in smaller bubbles.
As the thinning of the soap film progresses, the bubble becomes unstable and ruptures.
The effect of viscoelasticity has been pioneered by de Gennes \cite{deGennes:1996}, who refined the energy balance argument of the classic Taylor-Culick solution \cite{Culick:1960tl} and by Debr\'{e}geas {\it et al.} \cite{Debregeas:1995tg, Debregeas:1998}, who studied the bubble's stability in relation to their size.

In this work, we take a completely different approach.
We assume that the rupture of foam bubbles is a statistical event in which the probability depends on the mechanical properties of the fluid only implicitly.
The implicit dependence may be estimated by using statistical inference of the observed probability.
Formally, we assume that the probability of rupturing a bubble of size $r$ follows the power law such that
\begin{equation}
    p(r)\propto r^k,
\end{equation}
where the exponent $k$ carries the system's physical properties.

For experimental determination of $k$, we use the time-lapse photography technique.
Bubble foam decay is a slow process that can take several hours in some settings.
We set up a slow video imaging system, commonly referred to as a `time-lapse' in the photograph industry, to observe the time evolution of bubble foam for a long period of time up to several hours.
By counting the number of bubbles in each image, we obtain the size distribution of the bubbles.
With further analysis of bubble size statistics, the power exponent $k$ is indirectly measured.

We note that this approach is partly based on the soap-film network idea of Beenakker. \cite{Beenakker:1986, Beenakker:1987}
According to Beenarkker's model, soap foam is considered as a network of bubbles, and rupture is a process of losing a network node. 
In this perspective, the stability of bubble foam is a network properties that is connected to deeper topics such as topological disorder.
Even though we do not claim that our result is connected to such a deeper topic of statistical physics, our approach may provide a different perspective to the physical process of foam decaying.

\section{Experimental Method}\label{sec2}

\subsection{Setup and Procedure}

\begin{figure}
\begin{centering}
\includegraphics[width=7.5cm]{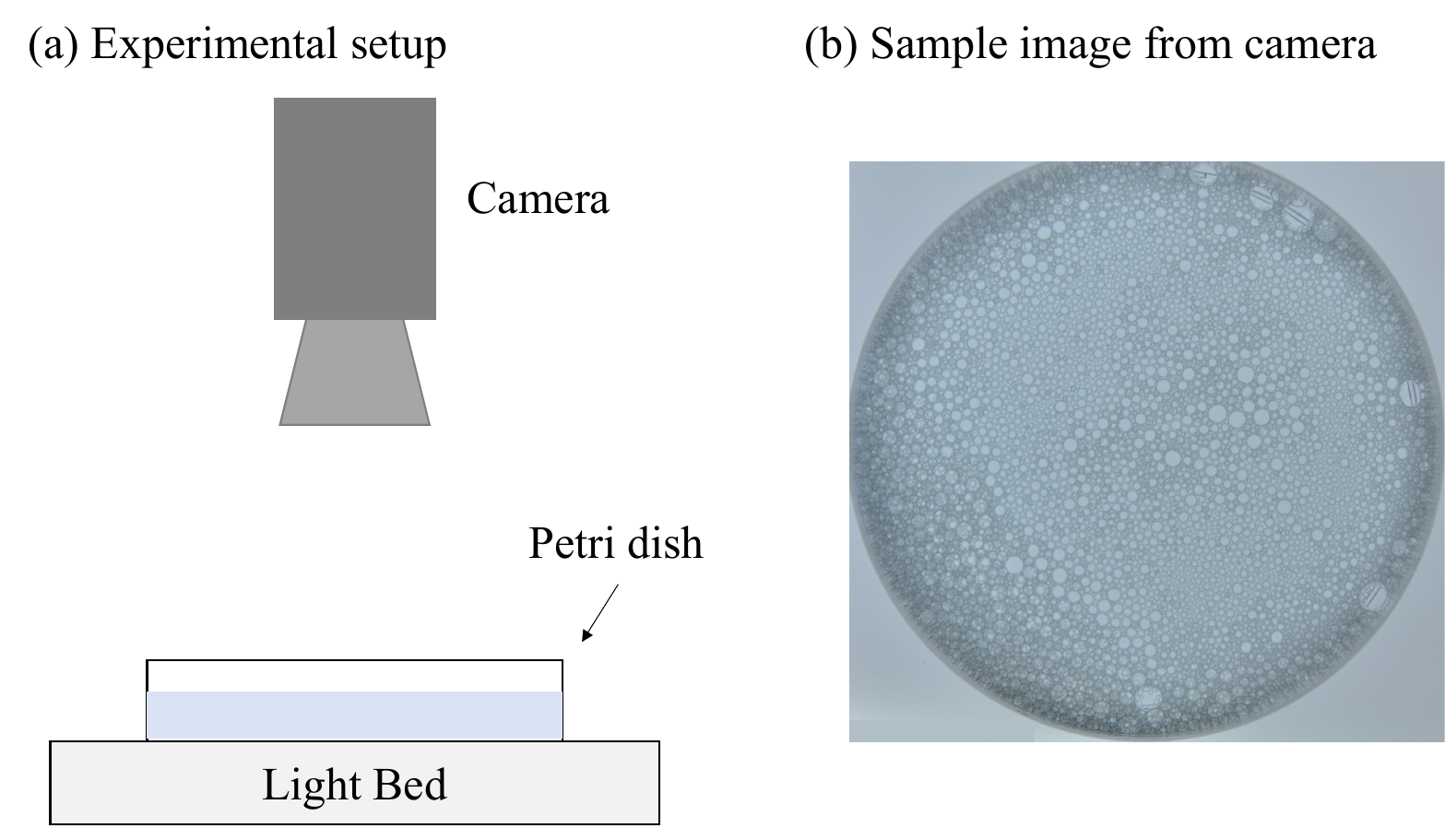}
\par
\end{centering}
\caption{
(a) A schematic diagram of our experimental setup.
(b) A typical image taken from the camera.
\label{fig:apparatus}}
\end{figure}

The experimental setup is extremely simple, as shown in Fig. \ref{fig:apparatus}(a).
The apparatus consists of only one main components, a Petri dish.
The diameter and depth of the petri dish is 8 cm and 0.5 cm, respectively, and therefore the petri dish holds the volume of approximately 32 $\rm cm^3$.

First, we half-fill the Petri dish by pouring 16 $\rm cm^3$ of soap solution.
Our soap solution is a mixture of distilled water and commercial dish soap (Dawn Ultra, Proctor \& Gamble), and we prepare the solution of different concentrations. 
We define concentration as a fraction of the mass of liquid soap to the total mass, that is, $c_0=(m_s)/(m_s+m_w)$, where $m_s$ and $m_w$ are the mass of liquid soap and distilled water used in the production of the soap solution, respectively.
The current experiment runs includes data from $c_0$=0.001, 0.002, 0.005, 0.01, 0.02, 0.05, 0.1 (0.1\% to 10\%).
As summarized in Table \ref{tab:exp_settings}, the measurements are repeated twice for each $c_0$.

For the reader's information, 1\% in the mass fraction is equivalent to 0.96\% in the volume fraction, as the density of liquid soap is $1.04\pm0.01 \, \rm g/cm^{3}$ according to our in-house measurement.
In literature concerning soap films, the volume fraction is reported more frequently, but in this paper we use the mass fraction because it is easier to measure mass than volume when the sample size is relatively small.
We also note that in the past, namely before 2010, P\&G's Dawn `Non-Ultra' was used more frequently for soap film research in the US.
However, as the supply of this `Non-Ultra' model has been cut in the US retail market in mid-2000, the use of `Ultra' model became more popular.
The Ultra model is claimed to be thrice as concentrated as the original Non-Ultra model.
According to the safety date sheet published by the manufacturer, the Ultra model is 10 to 15 percent solution (by mass) of sodium alkyl sulfate (CAS\# 68585-47-7, syn. sulfuric acid, mono-C10-16-alkyl esters, sodium salts), which is an industrial way of referring to sodium dodecyl sulfate with impurities (SDS, sodium alkyl sulfate with 12 carbons).
Considering that the molar mass of SDS is 288, the Ultra model can be considered as a solution of 323 mM of SDS.
Using the widely referred critical micelle concentration (CMC) of SDS of 8.2 mM, we estimate that the SDS' concentration in Dawn Ultra is in the range between 40 to 60 times of CMC.
Therefore, the 2\% soap solution is closest to the CMC in terms of SDS concentration.
However, this estimate should only be considered as a rough guide, because the commercial dishsoap is a mixture of many surfactants, and such a case the CMC is poorly defined.

Second, once the soap solution is poured onto the Petri dish, we create bubbles by stirring the liquid using a mesh.
The mesh consists of 20 horizontal and 20 vertical wires, producing 361 small rectangles of size 0.2 cm $\times$ 0.2 cm. 
The mesh is inserted into and extracted from the soap solution in the Petri dish.
We repeat the insertion and extraction of the mesh 50 times in 50 seconds, and then the Petri dish is covered by roughly one layer of soap bubbles.

Third, the bubble foams produced are left to collapse by themselves, and they are monitored by a camera that is mounted directly above them. 
We use a digital single lens reflector camera (Nikon D90), controlled remotely using a commercial software (ControlMyNikon).
Using the time-lapse functionality, the camera captures the bubbles on the Petri dish at 1/10 Hz to 1/50 Hz up to 5 hours.
The interval and total duration differ by the experimental runs, most importantly by $c_0$, as summarized in Table \ref{tab:exp_settings}.

\begin{table}
\begin{centering}
\begin{tabular}{c|c c c c}
\hline 
No. & $c_0$ & Interval & Duration & Total frames \\
 &  & (s) & (hr) &  \\
\hline 
\hline 
1 &\, 10\% & 50 & 5 & 360 \\
\hline 
2 &\, 10\% & 50 & 4 & 288 \\
\hline 
3 &\, 5\% & 50 & 4 & 288 \\
\hline 
4 &\, 5\% & 50 & 4 & 288 \\
\hline 
5 &\, 2\% & 25 & 4 & 576 \\
\hline 
6 &\, 2\% & 25 & 3 & 432 \\
\hline 
7 &\, 1\% & 25 & 3 & 432 \\
\hline 
8 &\, 1\% & 25 & 3 & 432 \\
\hline 
9 &\, 0.5\% & 50 & 5 & 360 \\
\hline 
10 &\, 0.5\% & 50 & 5 & 360 \\
\hline 
11 &\, 0.2\% & 50 & 5 & 360 \\
\hline 
12 &\, 0.2\% & 25 & 3 & 432 \\
\hline 
13 &\, 0.11\% & 25 & 3 & 432 \\
\hline 
14 &\, 0.10\% & 10 & 2 & 720 \\
\hline 
\end{tabular}
\par\end{centering}
\caption{List of experimental conditions.
\label{tab:exp_settings}}
\end{table}

\subsection{Data Analysis}

Each experimental run produces 288 to 700 sequential images like an example presented in Fig. \ref{fig:apparatus}(b).
Using in-house software using MATLAB (Mathworks), we extract the locations and radii of bubbles in each image frame. 
To ensure the integrity of software-based bubble counting, we manually counted bubbles in several random frames, and the difference between automated and manual measurements was less than 5\%.

Once the measurement is complete, the bubble radii are statistically analyzed.
In the current study, the  bubble locations are only used for sanity check and error detection.

We define the bubble radius as a random variable $X$, and in each image frame, the values of the random variable represent the radii of bubbles observed in that specific frame.
Formally,
\begin{equation}
X_{m}^{(\alpha)}=\{x_{m,n}^{(\alpha)}\}=\{x_1, x_2, \ldots, x_N\},
\end{equation}
where $\alpha$ represents the experimental condition (run no. in Table \ref{tab:exp_settings}), $m$ is the frame number, and $n=1,\cdots,N$ is an index to bubbles in the frame $m$. 
Here, $N$ represents the total number of bubbles in the frame $m$, which obviously depends on $\alpha$ and $m$.
Repeating the analysis for all frames, we get the {\it transient} size distribution of the bubbles.
We note that $m$ is used interchangeably with time because $t=mt_0$, where $t_0$ is the time interval between frames.
We also note that the indices may be omitted if there is no possibility of confusion.

After the initial extraction of bubble radii using the software, we apply post-processing.
In the preliminary analysis, we find that our in-house software is unable to make a reliable measurement of bubble radii smaller than 0.15 mm. 
Therefore, in the post-process we simply remove all the bubbles smaller than 0.15 mm, and this process defines the lower bound of our size distribution.

Before we exit this section, we note that the experiment is carried out in an open chamber where the humidity was monitored but not deliberately controlled.
We monitored relative humidity using hygrometers (Govee H5074).
The precision of this model is estimated to be around 3\%p, as provided by the manufacturer.
However, we installed two units side by side, and the two readings are consistent within 1\%p.
Taking the average value, we estimate that our experimental runs are carried out under a relative humidity of 34$\pm5$\%.

\section{Result and Discussions}\label{sec3}

\begin{figure*}
\begin{centering}
\includegraphics[viewport=25 0 690 580, width=11cm]{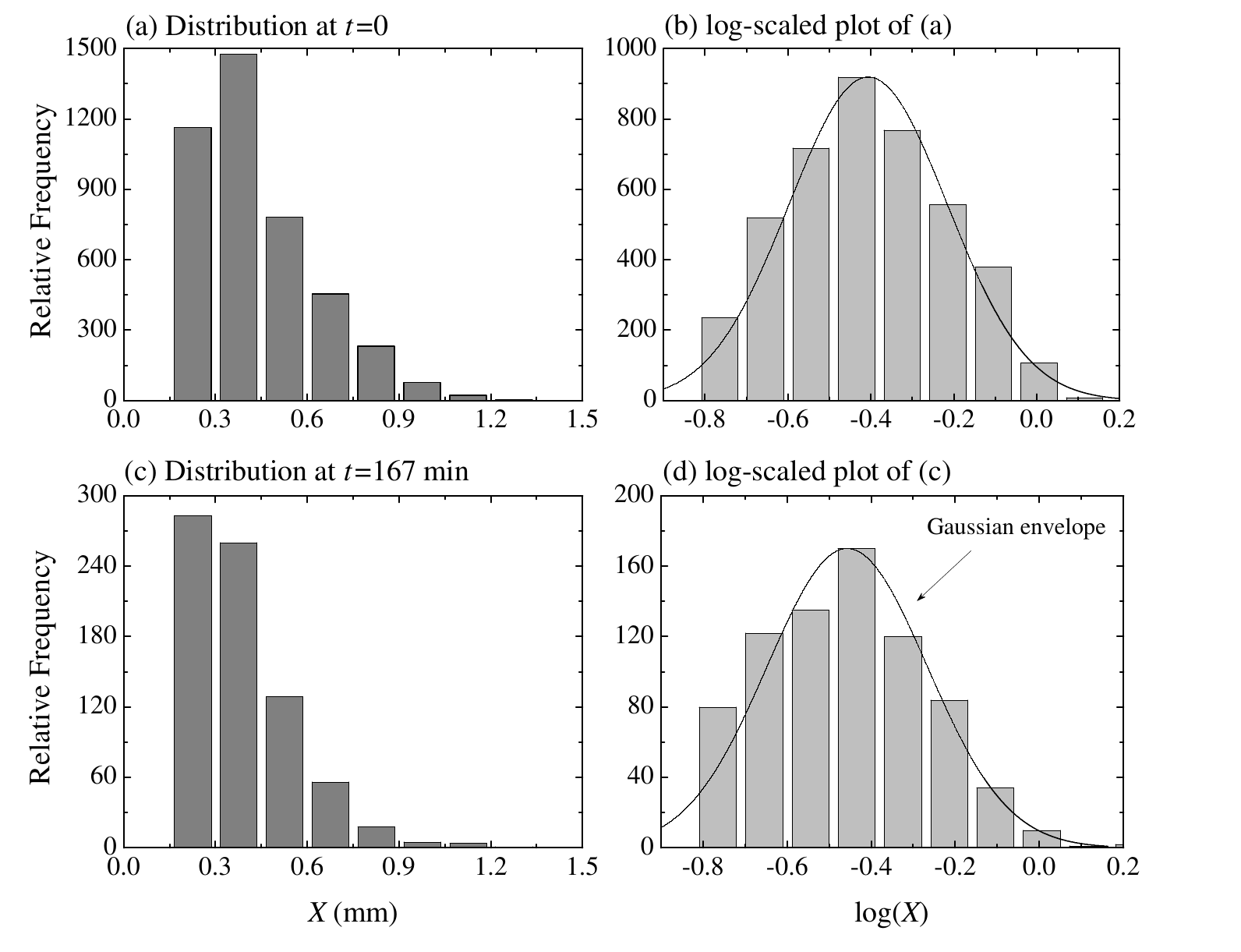}
\par
\end{centering}
\caption{
Size distribution of bubble radii $X$ at (a) $t=0$ and (c) $t=10^4$ s (167 min).
The distributions of $\log X$ are plotted in (b) and (d).
At $t=0$, the distribution of $\log X$ fits well under the Gaussian envelope (solid curve), indicating that the distribution is log-normal.
However, as time elapses, the distribution deviates from the log-normal.
\label{fig:distribution}}
\end{figure*}

In Fig. \ref{fig:distribution}, typical distributions of the bubble radius $X$.
These two probability distributions in Figs. \ref{fig:distribution}(a) and \ref{fig:distribution}(c) are obtained from the experiment using the 10\% soap solution (the experimental run no. 2 in Table \ref{tab:exp_settings}), at two different times: $t=0$ in (a) and $t=10^4$ s (167 min) in (c).
We note that these distributions are archetypal and share common features with results from other experimental runs, as will be detailed in the following paragraphs.

\subsection{Initial Log-normality}

We find that the initial distribution of $X$ at $t=0$ is considered to be roughly log-normal.
As shown in Fig. \ref{fig:distribution}(a), the initial distribution of bubble radii have a clear mode near $x=0.3$ mm in this specific case.
Using the data set, we can simply calculate a few representative values: the mean is 0.432 mm, the median is 0.388 mm, the variance (second moment about the mean) is measured $(0.196\,\rm mm)^2$, and the skewness (third moment about the mean) is 1.08 $\rm mm^3$.
These numbers show that the distribution is positively skewed, as obvious as in Fig. \ref{fig:distribution}(a).

However, if we plot the distribution of $\log X$, the distribution changes to roughly Gaussian.
As shown in Fig. \ref{fig:distribution}(b), the distribution is roughly symmetric about the mode and is well capped by the Gaussian envelope.
These characteristics suggest that $\log X$ is normally distributed and that the distribution of $X$ is log-normal.

We note that the log-normal distribution has been observed in fragmentation problems where an initial large volume of fluid undergoes a cascade of breakup processes \cite{Villermaux:2007wa}. 
In the physical process of preparation of the initial bubble layer, we also apply external forces on the bubbles in a repetitive manner, and this is perhaps the reason why we observe the log-normality.

\subsection{Time evolution}

However, the initial log-normality changes over time as bubbles decay, and most importantly, there is a noticeable trend that larger bubbles decay faster than smaller ones. 
In Fig. \ref{fig:distribution}(c), the distribution of bubbles at $t=10^4$ s (167 min) is presented. 
At this time, the mean, median and variance are 0.386 mm, 0.345 mm and $(0.187\,\rm mm)^2$, respectively.
Compared to the initial distribution, both median and mean have decreased, indicating that larger bubbles decay faster. 
The skewness increases to 1.92 $\rm mm^3$, showing a strong disparity in the bubble decay rate up to their size.
The distribution of $\log{X}$, as seen in Fig. \ref{fig:distribution}(d), substantially deviates from the Gaussian envelope, and the distribution is no longer considered log-normal.

\begin{figure*}
\begin{centering}
\includegraphics[viewport=25 25 750 735, width=11cm]{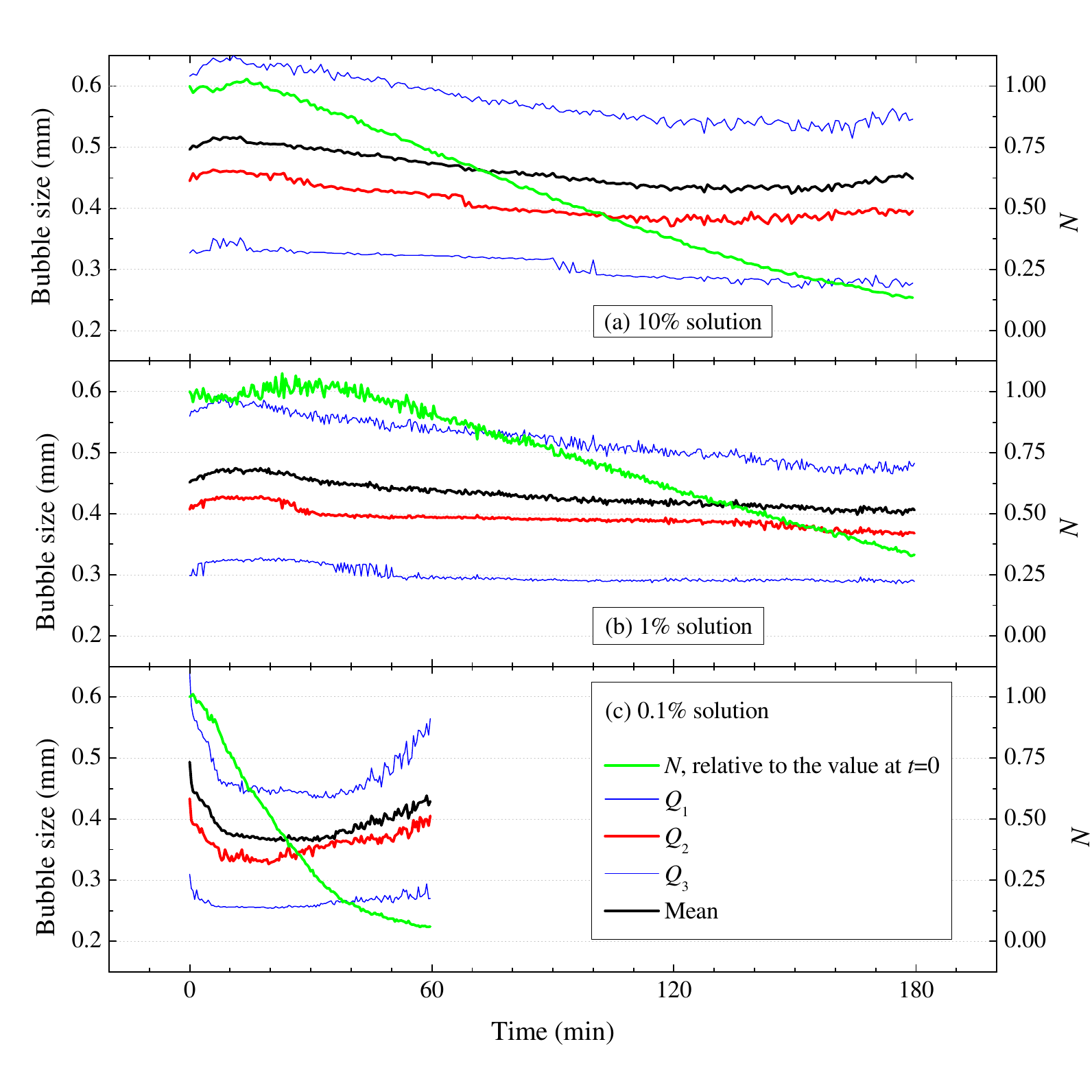}
\par
\end{centering}
\caption{
Representative values of the distribution of $X$ plotted with respect to time. 
In time, the mean and the median tends to decreases, until $N\simeq0.3N_0$.
\label{fig:mean}}
\end{figure*}

The size dependence of bubble decay is also illustrated in Fig. \ref{fig:mean}.
In the figure, we plot representative values, the mean ($\bar{x}$), first quartile ($Q_1$), median ($Q_2$) and third quartile ($Q_3$), as well as the total number of bubbles ($N$), with respect to time. 
We selected three experimental runs for comparison; run no. 2 (10\% soap solution), no. 8 (1\% soap solution) and no. 13 (0.1\% soap solution).
In all three cases, $Q_1$ is seen to decrease faster than $Q_3$.
This observation indicates that the larger bubble decays faster than the smaller bubbles, regardless of the soap concentration.

\subsection{Size Dependence of Decay Rate}

Figure \ref{fig:mean} shows not only that larger bubbles are less stable than smaller bubbles, but also that the stability of bubbles strongly depends on the soap concentration they are initially made of. 
For example, when bubbles are made of 10\% soap solution, half of the bubbles disappear after 100 min. 
However, when the bubbles are made of 0.1\% soap solution, it took only 20 min.
In contrast, bubbles made of the 1\% soap solution are more stable, as half of them remain intact after two hours.
This suggests that bubbles are not stable when the soap concentration is too high or too low, but they must be an optimal concentration at which bubbles are most stable.

For quantitative measurement of the decay rates of bubbles, we calculate the relaxation time scales of the decay for each run. 
We assume that bubbles are spherical, and then we define the gross bubble volume $V$ as follows:
\begin{equation}
V^{(\alpha)}(mt_0)=  \frac{4}{3}\pi \sum_n \left[x_{m,n}^{(\alpha)}\right]^3.
\end{equation}
From the analyses, we find that $V_t$ decays in an exponential manner, following the form 
\begin{equation}
V(t)=V_{0} e^{-t/\tau},
\label{eq:exp_decay}
\end{equation}
where $V_{0}$ is the value of $V(t)$ at its maximum, and $\tau$ is a relaxation time scale.

We then separate the data set into two subgroups; one is the large bubble group $Y$ and the other is the small bubble group $Z$. 
Formally, they are defined as follows:
\begin{eqnarray}
Y_{m}^{(\alpha)}&=&\{y_n \in X_{m}^{(\alpha)} | y_n>0.70 \,{\rm mm} \}, {\rm\, and}\\
Z_{m}^{(\alpha)}&=&\{z_n \in X_{m}^{(\alpha)} | z_n<0.35 \,{\rm mm} \},
\end{eqnarray}
where we choose 0.70 mm and 0.35 mm because they are approximately the first and third quantile values at $t=0$.
Then, we can define the total volume of large bubbles and small bubbles, as well as their respective relaxation times, in a similar manner as used in Eq. \eqref{eq:exp_decay}.

\begin{figure}
\begin{centering}
\includegraphics[viewport=50 40 665 540, width=7cm]{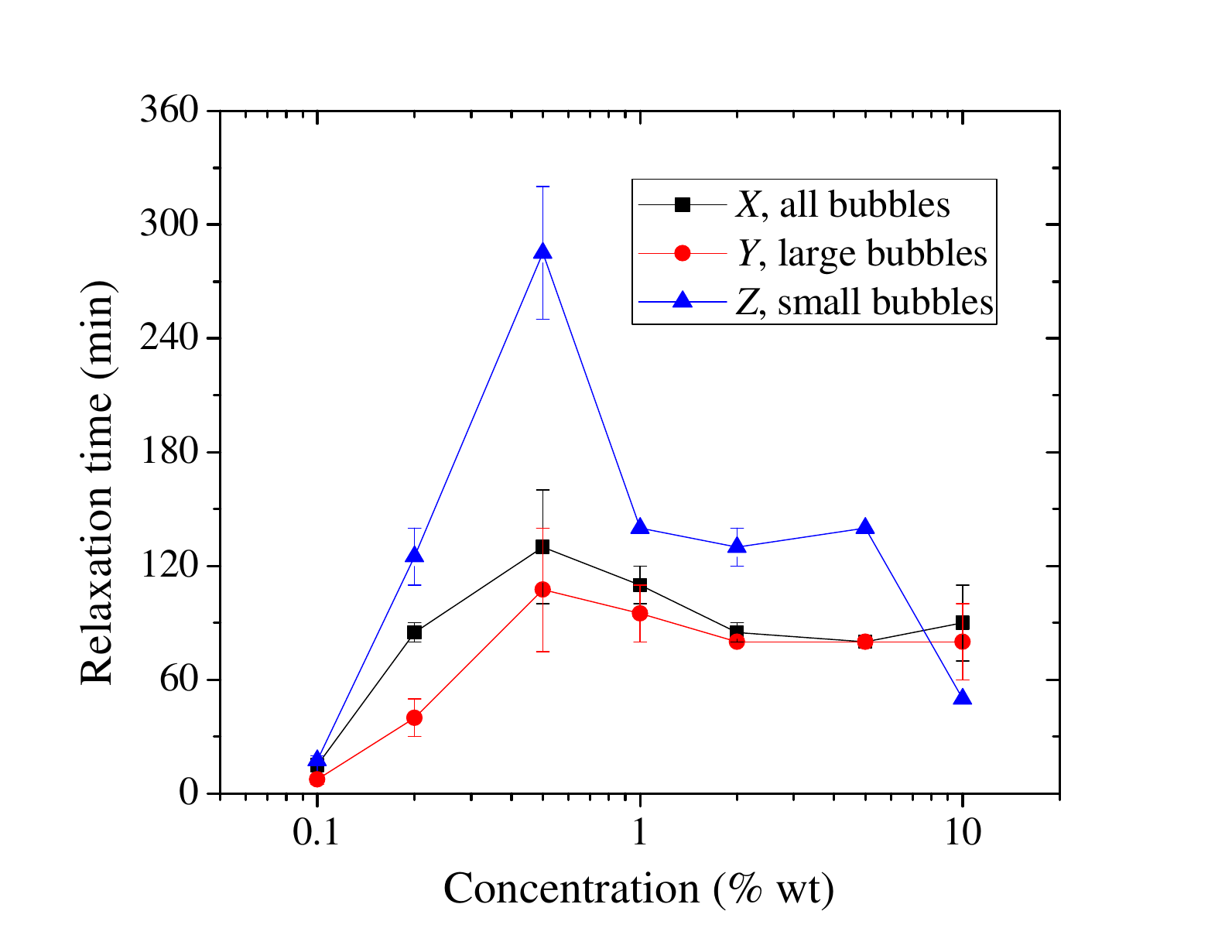}
\par
\end{centering}
\caption{
The volume of all bubbles, the volume of large bubbles ($>$ 0.70 mm), the volume of small bubbles ($<$ 0.35 mm), and the mean and median of bubbles.
\label{fig:relaxation}}
\end{figure}

In Fig. \ref{fig:relaxation}, we present the measurement of relaxation times using data in three subgroups, $X$, $Y$ and $Z$, which we call $\tau$, $\tau_\ell$, and $\tau_s$, respectively.
We first find that $\tau_\ell$ is substantially larger than $\tau_s$, confirming our observation from Fig. \ref{fig:distribution} that smaller bubbles decay much slower than larger bubbles.
Although relaxation times vary by soap concentration, the overall trend is clear; while $\tau_\ell$ ranges from 15 to 80 minutes, $\tau_s$ is much longer, exceeding 120 minutes on average.
At $c_0=0.5\%$, $\tau_s$ reaches the highest value, which means that 36.7\% of the smaller bubbles still remain in the Petri dish after 285 minutes (4 hours and 45 minutes).

We speculate that soap bubbles are most stable at $c_0=0.5$\% because at this concentration the Gibbs elasticity of the soap film is the greatest.
In a previous measurement using the same detergent, Sane {\it et al.} \cite{Sane:2018uo} found that 0.5\% soap solution makes the soap film with greater elasticity than the 1\% and 2\% solutions.
In the same work, they also found that the thinner film has greater elasticity at the same concentration; this might be the reason why smaller bubbles are more stable than larger ones.

\subsection{The Modeling of Data}

\begin{figure}
\begin{centering}
\includegraphics[width=7.4cm]{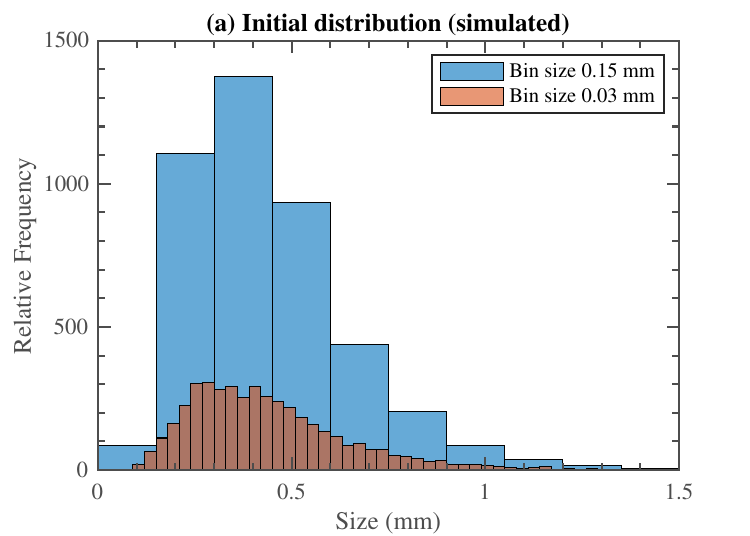}
\includegraphics[width=7.4cm]{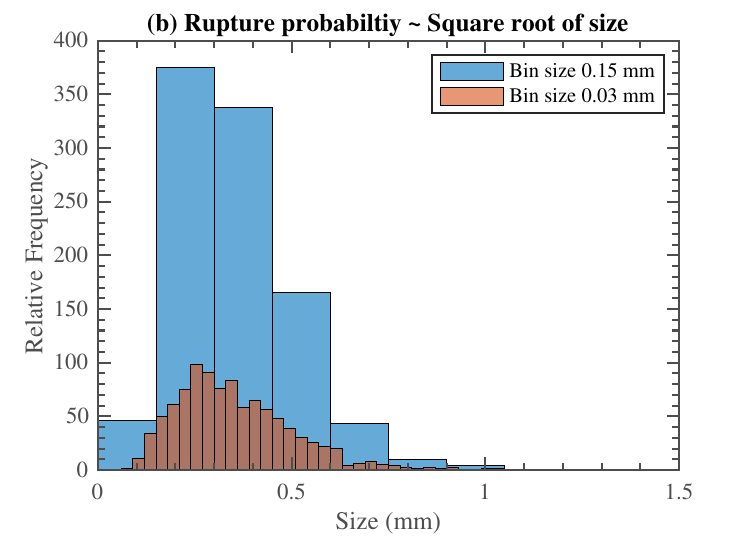}
\includegraphics[width=7.4cm]{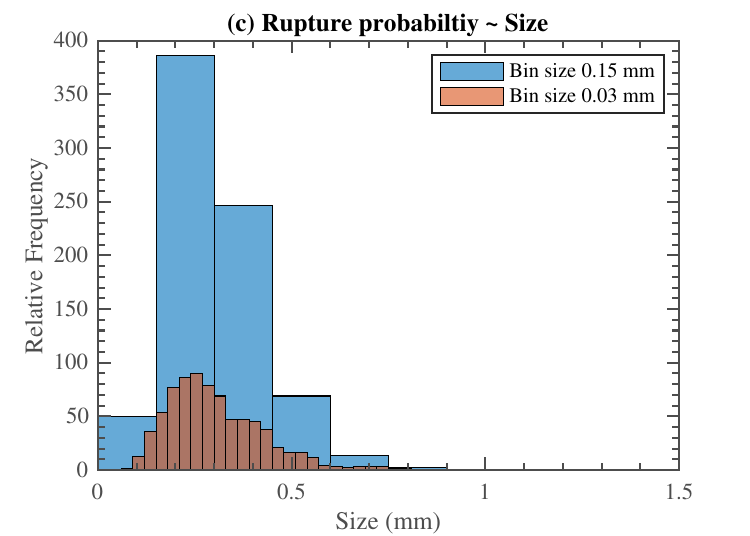}
\par
\end{centering}
\caption{
The simulated volume of all bubbles, the volume of large bubbles ($>$ 0.70 mm), the volume of small bubbles ($<$ 0.35 mm), and the mean and median of bubbles.
(a) Log-normal distribution at $t=0$.
(b) The time evolution of bubble size assuming $k=0.5$.
(b) The time evolution of bubble size assuming $k=1$.
\label{fig:model_calc}}
\end{figure}

For further understanding of the experimental observation, we make a simple toy model in which we assume that bubbles rupture in a probabilistic manner.
First, we generate an initial distribution to follow the log-normal distribution as shown in Fig. \ref{fig:model_calc}(a).
Second, each bubble will be determined to stay or disappear based on the probabilistic manner based on the power law.
We assume that the probability of rupture is expressed as 
\begin{equation}
    p_r\propto x^k,
\end{equation}
where $k$ is an exponent.
Third, we iterate the second step for a certain amount of time using Monte Carlo simulation.

We find that our Monte Carlo simulation matches with our experiments most when $k=0.5$.
In Fig. \ref{fig:model_calc}(b), the simulated distribution using $k=0.5$ is shown, and its level of similarity with our data in Fig. \ref{fig:distribution}(c) is astonishing. 
We note that if a different value of $k$ is used, the distribution does not look like the experiment. 
For example, if $k=1$ is used as seen in Fig. \ref{fig:model_calc}(c), the larger bubbles are more likely to rupture, leading to a less positively skewed distribution than the experimental observation in Fig. \ref{fig:distribution}(c).

Our result is consistent with the experimental observation of Debr\'{e}geas {\it et al.} \cite{Debregeas:1995tg, Debregeas:1998} in that the larger bubbles are less stable than the smaller bubbles. 
However, their conclusion that the decay rate is proportional to the bubble size is not consistent with ours.
While we cannot completely rule out the possibility that $k=1$, it seems $k=0.5$ fits the experimental data most well.

\section{Summary}\label{sec4}

In summary, we investigated the bubble size distribution in soap foam. 
Using a commercial dishsoap, we produced a thin layer of bubbles in a Petri dish, and using a slow video imaging technique, we observed the time-dependent bubble size distribution.

We find that the bubble size distribution is initially log-normal.
However, as time elapses, the larger bubbles decay much faster than smaller bubbles, which makes the distribution deviate from the log-normal distribution.
When the soap concentration is neither too small nor too large, i.e. between 0.2\% to 5\%, the stability of bubbles depends on their size; the larger bubbles (radii greater than 0.70 mm) have a decay rate at least three times higher than their smaller brothers (radii less than 0.35 mm).
Our investigation suggests that the probability of rupture is proportional to the 1/2 power to the bubble size.

This study was motivated by the perspective that soap foam is considered as a network.
In this network, bubble rupture is a process in which a pair of nodes are merged randomly, and this random process is driven by the physical properties of the fluid.
The extraction of such physical information requires further investigation.

\section*{Acknowledgement}

This work is supported by Konkuk University in 2023.

\bibliography{bubble2018}

\end{document}